\documentclass[aps,prd,reprint,floatfix,longbibliography,nofootinbib]{revtex4-1}
\usepackage[utf8x]{inputenc}
\usepackage{microtype}
\usepackage{graphicx}
\usepackage[dvipsnames]{xcolor}
\usepackage{rotating}

\usepackage{amsmath}
\usepackage{amssymb}
\usepackage{amsfonts}

\usepackage{tabu}
\usepackage{booktabs}
\usepackage{multirow}
\usepackage{array}
\usepackage{enumitem}
\usepackage{url}
\usepackage[colorlinks=true,urlcolor=blue,linkcolor=blue,citecolor=blue]{hyperref}
\usepackage{braket}

\def\0#1#2{\frac{#1}{#2}}

\def\s0#1#2{\mbox{\small{$ \frac{#1}{#2} $}}}

\AtBeginDocument{%
	\newwrite\bibnotes
	\def\bibnotesext{Notes.bib}
	\immediate\openout\bibnotes=\jobname\bibnotesext
	\immediate\write\bibnotes{@CONTROL{REVTEX41Control}}
	\immediate\write\bibnotes{@CONTROL{%
			apsrev41Control,author="08",editor="1",pages="1",title="0",year="1"}}
	\if@filesw
	\immediate\write\@auxout{\string\citation{apsrev41Control}}%
	\fi
}%

\begin{document}

\title{The Hawking Energy on the Past Lightcone in Cosmology}

\author{Dennis Stock}
\affiliation{University of Bremen, Center of Applied Space Technology and 
	Microgravity (ZARM), 28359 Bremen, Germany}

\begin{abstract}
 This work studies the Hawking energy in a cosmological context. The past lightcone of a point in spacetime is the natural geometric structure closely linked to cosmological observations. By slicing the past lightcone into a 1-parameter family of spacelike 2-surfaces, the evolution of the Hawking energy down the lightcone is studied. Strong gravitational fields may generate lightcone self-intersections and wave front singularities. We show that in the presence of swallow-tail type singularities, the Hawking energy and its variation along the null generators of the lightcone remains well-defined and subsequently discuss its positivity and monotonicity.

\end{abstract}

\maketitle

\section{Introduction}
\label{sec:intro}
A general and consistent notion of energy or mass poses a difficulty in the context of general relativity. Due to the weak equivalence principle, the energy momentum distribution of the gravitational field is locally vanishing for a freely-falling observer moving along a geodesic. Because of these local considerations, quasilocal constructions were brought forward, see for instance \cite{Szabados:2004vb} and references therein. Amongst the candidates is a construction by Hawking \cite{Hawking:1968qt}, which will be referred to as Hawking energy in the following. Based on a closed spacelike 2-surface $S$ in spacetime, it phenomenologically aims to relate the energy/matter content enclosed by $S$ to the amount of light bending on $S$. 

A sensible energy definition should match previously established concepts of energy in highly symmetric or asymptotic settings, such as the ADM- or Bondi-mass, but still be general enough to also apply to more general set-ups. A central challenge for many quasilocal energy definitions is to provide a sensible meaning to the concepts of positivity and monotonicity in a physically realistic and general enough context. For asymptotically flat spacetimes filled with matter obeying the dominant energy condition (DEC), several positive mass theorems for the ADM-mass \cite{Schon:1979rg,schoen1981,witten1981} as well as for the Bondi-mass \cite{Ludvigsen_1981,Horowitz:1981uw,PhysRevLett.48.369} were established. Later, positivity of mass was extended also to asymptotically AdS-spacetimes and to Einstein-Maxwell theory \cite{gibbons1983}. Closely linked to the question of positivity is the Penrose conjecture \cite{Penrose:1973um,Mars:2009cj}, relating the total mass of a spacetime to the area of the outermost apparent horizon. The corresponding Riemannian version, the Riemann-Penrose inequality, was proven by Huisken \& Ilmanen \cite{huisken2001} using the observation by Geroch that the Hawking energy behaves monotonously under the inverse mean curvature flow \cite{1973NYASA.224..108G}; it was also independently proven by Bray \cite{bray2001}. More generally, in \cite{Bray:2006pz} the authors examined under which flows the Hawking energy is monotonous.

The context in which the present work studies energy in a spacetime is observational cosmology \cite{Ellis1985315,ellis_maartens_maccallum_2012}, whose fundamental objective is to infer properties of the universe solely based on local observations. In particular, we would like to find a meaningful notion of energy for the observable universe, that is, the part of spacetime causally connected to and in the past of the observation event. The natural geometric object directly related to the causal boundary, but also to cosmological observations, is the past lightcone of an observer.
We stress that the lightcone is a geometric object associated to the spacetime at a given point without any further specifications. Therefore, a natural question is whether a well-defined notion of energy on the past lightcone exists. Positivity of energy for null-geodesically complete, globally smooth lightcones was shown in \cite{Chrusciel:2014gja}. Specifically for the Hawking energy, positivity and monotonicity results were established in \cite{Christodoulou:1988,Eardley:1979dra}. However, due to strong gravitational fields, potentially sourced by local inhomogeneities, the lightcone might develop singularities and caustics \cite{Ellis:1998ha,Friedrich:1983vi,Hasse_1996,arnold1985singularities}, see also \cite{Perlick2004} for a general overview about lightcones in the context of gravitational lensing. The aim of this paper is to study the properties of the Hawking energy in such a cosmological set-up, even admitting certain types of singularities which generically appear in lightcones.\\

This work is structured as follows. The Hawking energy and its main properties are discussed in section \ref{sec:Hawkingenergy}, before the cosmological set-up together with the slicing construction of the lightcone is explained in section \ref{sec:cosmolsetup}. The weak lensing regime in absence of self-intersections is studied in section \ref{sec:weaklensing}. The effect of self-intersections on the lightcone geometry is addressed in section \ref{sec:stronglensing}. Section \ref{sec:singularities} establishes the well-definedness of the Hawking energy and its derivative in the presence of swallow-tail type singularities. The rescaling freedom of the null generators of the lightcone is discussed in section \ref{sec:rescaling}, before moving to a discussion on monotonicity and possible extensions in sections \ref{sec:monotonicity} \& \ref{sec:extensions}. We conclude in section \ref{sec:conclusions}.

\section{Hawking Energy}
\label{sec:Hawkingenergy}
Unless stated otherwise, we assume a globally hyperbolic Lorentzian spacetime $(M,g)$ satisfying the Einstein field equations (EFEs):
\begin{equation}
R_{ab}-\frac{1}{2}R g_{ab}=8\pi T_{ab}\quad,
\end{equation}
with Ricci tensor $R_{ab}$, Ricci scalar $R$, and the energy-momentum tensor $T_{ab}$ satisfying the DEC. A potential cosmological constant can be accommodated in $T_{ab}$ in the following discussions. The signature convention is $(-+++)$ and we use units in which $c=G=1$.\\

A spacelike 2-surface $S$ in $M$ uniquely defines two distinct orthogonal null congruences, both of which are either future or past directed, represented by two null vector fields $l^a$ and $n^a$. These congruences are often referred to as outgoing and ingoing, their expansion scalars are denoted by $\theta_+=\nabla_al^a$ and $\theta_-=\nabla_an^a$ respectively, where $\nabla_a$ is the covariant derivative associated with the spacetime metric $g$. Hawking's original definition \cite{Hawking:1968qt} of the energy $E(S)$ associated with a spacelike surface $S$ of spherical topology reads:
\begin{equation}
E(S):= \frac{\sqrt{A(S)}}{(4\pi)^{3/2}}\left(2\pi+\frac{1}{4}\int_S\theta_+\theta_-\,\mathrm{d}S\right)\quad,
\label{eq:energy}
\end{equation}
where $A(S)=\int_S \mathrm{d}S$ denotes the area of the surface $S$ given in terms of the pullback $\mathrm{d}S$ of the canonical spacetime volume form onto $S$. This definition satisfies several important limits briefly reviewed here, see also \cite{Szabados:2004vb,Eardley:1979dra}:
\begin{itemize}
	\item[(i)] The Hawking energy of any point in spacetime should vanish, hence $E(S)\rightarrow 0$ for $S$ degenerating to a point.
	\item[(ii)] For a small sphere of (area) radius $r\rightarrow 0$ about point $p$, one finds for the leading order in $r$ \cite{Horowitz:1982}:
	\begin{align}
	E(S)&\sim r^5\, B_{abcd}t^at^bt^ct^d\ge 0\quad\text{in vacuum}\\
	E(S)&\sim r^3\,T_{ab}t^at^b\quad\text{in non-vacuum}
	\end{align}
	with the Bel-Robinson tensor $B_{abcd}$ \footnote{The Bel-Robinson tensor is defined as\\ $B_{abcd}:=C_{aecf}\,C_{b\;d}^{\;e\;f}-\frac{3}{2} g_{a[b}C_{jk]cf} C^{jk\;f}_{\;\;\;d}$ .}, $t^a\in T_pM$ a unit timelike vector orthogonal to $S$, and the energy-momentum tensor $T_{ab}$. If the DEC holds, then $E(S)\ge0$ also in the non-vacuum case.
	\item[(iii)] For large spheres near null infinity $\mathcal{I}^\pm$, the Bondi-Sachs energy is recovered \cite{Hawking:1968qt}: $E(S)\rightarrow E_{\text{Bondi-Sachs}}$.
	\item[(iv)] For large spheres near spatial infinity $i^0$, the ADM-mass is recovered \cite{Szabados:2004vb}: $E(S)\rightarrow E_{\text{ADM}}$ .
	\item[(v)] In a spherically symmetric spacetime, the Hawking energy coincides with the Misner-Sharp energy, e.g.\ \cite{Carrera:2008pi}.
	\item[(vi)] If $S$ is a metric sphere in Minkowski spacetime: $E(S)=0$ \cite{Szabados:2004vb}.
	\item[(vii)] Given a null hypersurface with $\theta_+=0$, for instance a non-expanding horizon or a Killing horizon. For any spacelike spherical cross section $S$, one finds:
	\begin{equation}
	E(S)=\sqrt{\frac{A(S)}{16\pi}}\quad.
	\end{equation}
	In particular, for a cross section of the event horizon of a Kerr-Newman black hole, the irreducible mass $M_{\text{irr}}$ is recovered, see e.g.\  \cite{Eardley:1979dra}.
\end{itemize}
Two other properties one would expect from an energy definition are positivity and monotonicity. However, it appears that this is not given in the general case. Concerning positivity, it is worth pointing out that (vi) only holds for metric spheres and not for arbitrary topological spheres on Minkowski spacetime. In fact, the Hawking energy might become negative for suitably shaped spheres\footnote{In general, the Hawking energy turns negative if according to (\ref{eq:ww}) the mean curvature $H$ of $S$ within the spacelike hypersurface $\Sigma$ is large enough compared to the mean curvature $\tau$ of $\Sigma$ in $M$.}. In order to maintain a vanishing energy for any spacelike topological sphere in Minkowski space, Hayward proposed a modification by including shear and twist terms \cite{Hayward:1993ph}. However, it is negative for small spheres in vacuum \cite{Bergqvist_1994}. A general positivity result for maximal slices was obtained in \cite{Christodoulou:1988}. Furthermore, one would naturally expect the energy to increase if the domain, i.e.\ the surface $S$, is enlarged. Since in general there are many ways to enlarge $S$, one would have to specify a particular construction to give a more precise meaning to the statement. Eardley was able to construct a special family of surfaces along which the Hawking energy increases monotonously \cite{Eardley:1979dra}. This result is essential in order to establish monotonicity in the weak lensing case and will be discussed in greater detail in section \ref{sec:weaklensing}.

In the light of these results, a natural question is whether positivity and monotonicity of the Hawking energy can be established in particular, physically relevant set-ups, such as the past lightcone of an observer in cosmology.

\section{Cosmological Set-up}
\label{sec:cosmolsetup}
The cosmological context in which we aim to answer this question is provided by the observational approach by Ellis and others \cite{Ellis1985315,ellis_maartens_maccallum_2012}. Based solely on data on the past lightcone of an observer, it aims at deducing the spacetime geometry in the vicinity of the lightcone without further model assumptions. Mathematically, it constitues a characteristic final value problem, see e.g.\ \cite{ChoquetBruhat:2010ih,Chrusciel:2012ap} and references therein, with final data given on the past lightcone and a solution in the chronological past of the event is constructed by propagating the data on the lightcone into its interior via the EFEs. The observer is assumed to be a point $p$ in spacetime $M$ and a future-pointing normalised timelike vector $u^a\in T_pM$. This is a good approximation as long as the duration of observation is negligible compared to the dynamical timescale of the universe. Almost all cosmologically relevant information, such as light and gravitational waves, travels with the speed of light, hence the central geometric object of interest is the past lightcone $C^-(p)$ of the observer at $p\in M$. It is a null hypersurface and can be uniquely constructed once the point $p\in M$ is specified. In Minkowski spacetime, it is an undistorted cone with topology $\mathbb{R}\times S^2$. However, the presence of matter or other inhomogeneities will in general deform the lightcone. Two regimes can be distinguished:
\begin{itemize}
	\item \textit{Weak Lensing Regime}: the lightcone remains an embedded surface, but is weakly deformed, preserving the $\mathbb{R}\times S^2$ topology. Hence, no multiple images of the same source appear.
	\item \textit{Strong Lensing Regime}: the lightcone is strongly deformed and intersects itself. Changes in topology cause multiple imaging.
\end{itemize}

\begin{center}
\begin{figure}
	\includegraphics[width=0.4\textwidth]{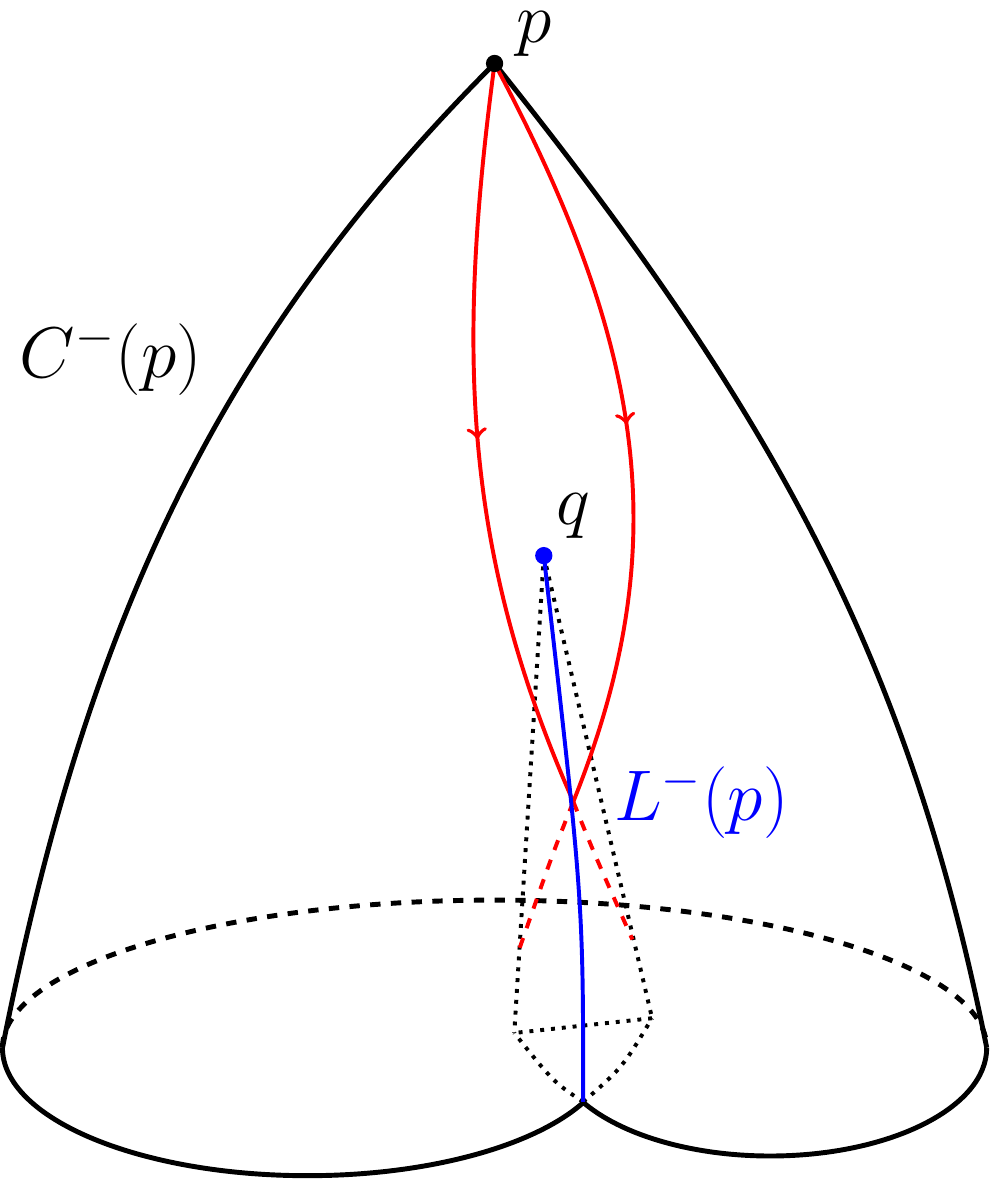}
	\caption{Lightcone $C^-(p)$ of point $p$ going through a gravitational lensing event causing $C^-(p)$ to intersect itself at the cut locus $L^-(p)$ (blue), which is part of the exterior $C^-(p)\cap\dot{I}^-(p)$. Two different null generators (red) intersect at the cut locus, after which they turn into the interior $I^-(p)$ (dashed). The conjugate point $q$, where infinitesimally close generators intersect, is of swallow-tail type. Two cusp ridges originating from $q$ remain in $I^-(p)$.}
	\label{fig:spherlens}
\end{figure}
\end{center}

More formally, the past lightcone $C^-(p)$ of a cosmological observer $(p,u^a)$ in a globally hyperbolic spacetime $M$ is the image of the exponential map $\exp_p$ along past-pointing null vectors $\in T_pM$ on its maximal domain of definition. Sufficiently close to $p$, the exponential map is always injective. At self-intersections, the exponential map fails to be injective, i.e.\ points may be reached along multiple null geodesics starting at $p$. Another crucial observation is that past null geodesics issued at $p$ are initially part of the boundary $\dot{I}^-(p)$ of the chronological past $I^-(p)$ of $p$, but might leave the boundary into the interior. Thus, they are not exclusively confined to $\dot{I}^-(p)$ but rather to $\dot{I}^-(p)\cup I^-(p)$. The last point along a null generator $\gamma(\tau)$ still in $\dot{I}^-(p)$ is called cut point of $\gamma$. The union of all cut points of all past-pointing null generators is then referred to as cut locus $L^-(p)$ of the past lightcone $C^-(p)$. Any point of a generator beyond the cut point lies in the chronological past of $p$ and therefore can also be reached along a timelike curve from $p$. At a cut point, multiple null generators intersect, either infinitesimally close generators resulting in a conjugate point, or globally different generators, see Fig.\ \ref{fig:spherlens}. Furthermore, since $\dot{I}^-(p)$ is an achronal boundary and therefore a Lipschitz continuous submanifold \cite{Hawking:1973uf}, the same holds for the part of the lightcone contained in the boundary, $C^-(p)\cap \dot{I}^-(p)$. Additionally, the cut locus has measure zero in $\dot{I}^-(p)$, thus, $C^-(p)\cap \dot{I}^-(p)$ is differentiable everywhere except at $p$ and the cut locus \cite{Perlick2004}.

In this cosmological set-up, we study the properties of the Hawking energy on the past lightcone $C^-(p)$ of a cosmological observer. In principle, the Hawking energy can be inferred directly from observations, at least within the ideal observational cosmology framework described in \cite{Ellis1985315}. For example, the energy associated with a round sphere of comoving radius $r$ in a FLRW-universe is $E=4/3\pi a^3r^3\rho$, where $\rho(t)$ is the matter density of the cosmic fluid and $a(t)$ the scale factor. 

In particular, we are interested in monotonicity properties of $E$ along a family of two dimensional slices $(S_t)$ down the lightcone. Before turning to more formal and rigorous statements in sections \ref{sec:weaklensing} \& \ref{sec:monotonicity}, we first provide an intuitive argument in favour of monotonicity.

\begin{center}
	\begin{figure}
		\includegraphics[width=0.45\textwidth]{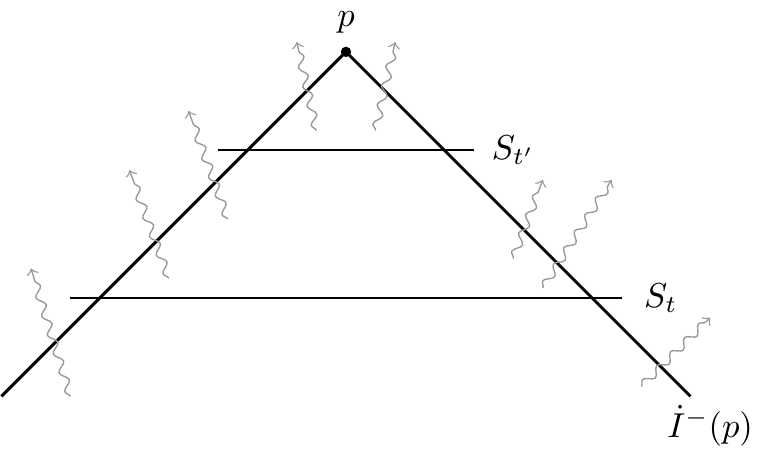}
		\caption{Causal matter can only leave $I^-(p)$ to the future, but nothing can enter from the outside. Thus, the Hawking energy should monotonously increase from $S_{t'}$ to $S_t$.}
		\label{fig:monotonicity}
	\end{figure}
\end{center}

The past lightcone in Fig.\ \ref{fig:spherlens} can be sliced into two dimensional spacelike surfaces, for instance by a one-parameter family of (partial) Cauchy surfaces $\Sigma_t$. The part of such a lightcone slice contained in the past causal boundary $\dot{I}^-(p)$ is denoted by $S_t$: $S_t:=C^-(p)\cap \dot{I}^-(p)\cap\Sigma_t$. Since $\dot{I}^-(p)$ is the past causal boundary of $I^-(p)$, any matter respecting the DEC can only leave $I^-(p)$ to the future, in particular, nothing can enter $I^-(p)$ from outside. Therefore, taking two different slices $S_t$ and $S_{t'}$ with $t<t'$ as depicted in Fig.\ \ref{fig:monotonicity}, matter may only leave $I^-(p)$ between $t$ and $t'$. Turning the argument around, the surfaces $S_t$ should enclose more and more matter towards the past. Each $S_t$ is typically a closed, spacelike surface and thus has an associated Hawking energy $E(S_t)$. By the above argument, the Hawking energy should then be monotonously increasing along the family $(S_t)$ down the lightcone. Though, this naive argument only holds for $\theta_+>0$ everywhere on $C^-(p)$ as we shall see later.

It is crucial to note that this argument only holds for surfaces which are part of the causal boundary. As mentioned above, the lightcone generators leave the boundary after self-intersections and the interior parts of $C^-(p)$, that is the part contained in the chronological past, can be penetrated by timelike curves. Therefore, we have to exclude the interior parts of $C^-(p)$ and restrict our monotonicity discussion to the part of the lightcone contained in the causal boundary $C^-(p)\cap \dot{I}^-(p)$. Also from a geometric point of view, $C^-(p)\cap \dot{I}^-(p)$ is a much better-behaved hypersurface than $C^-(p)$ since it is a Lipschitz manifold, whereas $C^-(p)$ might in general fail to be a manifold due to complicated self-intersections in the interior of $\dot{I}^-(p)$.\\

In the subsequent sections, we do not use Cauchy surfaces, but rather adopt the following construction in order to generate the 1-parameter family of lightcone slices. We start with an initial lightcone cut $S$ sufficiently close to $p$, guaranteeing that $S$ is a topological sphere and that the Hawking energy is positive due to the small sphere limit \cite{Horowitz:1982}. The part of the lightcone in the past of $S$ is generated by the past-pointing null geodesics associated with the null generators $l^a$. The lightcone can then be sliced into constant affine parameter distance slices $S_\lambda$, with $\lambda\geq 0$ and $S_{\lambda=0}=S$. However, since $l^a$ is a null vector field, we have a pointwise rescaling freedom $l^a\rightarrow\alpha l^a$, with $\alpha>0$ a function on $S$. After fixing the rescaling freedom in a suitable manner, as is done in the subsequent sections, we walk down a unit distance along the generators and arrive at a new spacelike cut, where the rescaling procedure is repeated etc. The function $\alpha$ extends to a function on $C^-(p)\cap \dot{I}^-(p)$ and encodes the particular choice of the lightcone foliation $(S_\lambda)$. Changing from one foliation with corresponding affine parameter $\lambda$ to another one with $\tilde{\lambda}$ relates the change of $E$ for each foliation via
\begin{equation}
\frac{\partial E}{\partial \tilde{\lambda}}=\frac{\partial E}{\partial \lambda}\frac{\partial\lambda}{\partial\tilde{\lambda}}\quad.
\label{eq:ooo}
\end{equation}
Thus, if $E$ is monotonously increasing along the $\lambda$-foliation, it increases along the new foliation if $\frac{\partial\lambda}{\partial\tilde{\lambda}}>0$, that is, if $\lambda$ is an increasing function of $\tilde{\lambda}$.

\section{Weak Lensing Regime}
\label{sec:weaklensing}
This section addresses monotonicity in scenarios excluding caustics, a discussion including these can be found in the next sections. The following results can be understood as an application of Eardley's findings \cite{Eardley:1979dra} to lightcones. Each lightcone slice on the boundary, $S_\lambda\subset C^-(p)\cap \dot{I}^-(p)$, comes with an associated Hawking energy $E(S_\lambda)$. In the following, it is assumed that each slice $S_\lambda$ is topologically a sphere, a brief discussion of different topologies can be found in section \ref{sec:extensions}. The change of the Hawking energy (\ref{eq:energy}) assigned to $S_\lambda$ along the outgoing null direction $l^a$ is given by $\partial_\lambda E(S_\lambda)\equiv \dot{E}(S_\lambda)$:
\begin{align}
\dot{E}(S_\lambda)&= \frac{E(S_\lambda)}{2A(S_\lambda)}\int_{S_\lambda} \theta_+\,\mathrm{d}S_\lambda+\nonumber\\
&+\frac{\sqrt{A(S_\lambda)}}{(4\pi)^{3/2}}\int_{S_\lambda}\left[ \dot{\theta}_+\theta_-+\theta_+\dot{\theta}_-+\theta_+^2\theta_-\right]\mathrm{d}S_\lambda\quad,
\label{eq:aa}
\end{align}
where we used $A(S)=\int_S\mathrm{d}S$ and $\dot{(\mathrm{d}S)}=\theta_+\mathrm{d}S$. At the same time, the vector field $l^a$ will be taken to be identical to the null generators of the past lightcone $C^-(p)$. Next, we will make use of the Sachs equation for the evolution of $\theta_+$ (see e.g.\ \cite{Hawking:1973uf}):
\begin{equation}
\dot{\theta}_+=-\frac{1}{2}\theta_+^2-\sigma_{ab}\sigma^{ab}-R_{ab}l^al^b\quad.\label{eq:sachseq}
\end{equation}
Since $l^a$ generates a null hypersurface, the vorticity term in the general Sachs equation is vanishing and thus absent in (\ref{eq:sachseq}). $\sigma_{ab}$ denotes the shear tensor of the congruence $l^a$. Using the EFEs, the right hand side of (\ref{eq:sachseq}) is non-positive if the DEC holds. 
The evolution equation of $\theta_-$ along $l^a$ can be derived from \cite{Gourgoulhon:2005ng}:
\begin{equation}
\dot{\theta}_-=D_a\Omega^a+\Omega_a\Omega^a-\frac{1}{2}{}^2R+\frac{1}{2}h^{ab}R_{ab}-\theta_+\theta_-\quad.\label{eq:-evolution}
\end{equation}
It describes the change of the expansion of the ingoing null congruence $n^a$ along the outgoing one. $\Omega_a=\nabla_ln_a$ denotes the change of the ingoing null vector along the outgoing one. $h_{ab}$ denotes the two dimensional Riemannian metric on $S_\lambda$ defined by the pullback of the spacetime metric $g_{ab}$ onto $S_\lambda$. They are related via 
\begin{equation}
h_{ab}=g_{ab}+l_an_b+n_al_b\quad.
\label{eq:hab}
\end{equation}
$D_a$ is the covariant derivative on $S_\lambda$ compatible with its induced metric $h_{ab}$ and related to the spacetime covariant derivative $\nabla$ via the projection operator onto $TS_\lambda$, $\Pi_a^{\;b}=\delta_a^{\;b}+l_an^b+n_al^b$. For instance, $D_aX^b=\Pi_a^{\;c} \Pi_d^{\;b}\nabla_cX^d$ for any $X^a\in TS_\lambda$. The two dimensional Ricci scalar of $S_\lambda$ is denoted by ${}^2R$. Using the EFE, we find
\begin{equation}
h^{ab}R_{ab}=R+2R_{ab}l^an^b=16\pi T_{ab}l^a n^b\ge0\quad,
\label{eq:DEC}
\end{equation}
if the DEC is satisfied. Inserting (\ref{eq:sachseq}), (\ref{eq:-evolution}), and (\ref{eq:DEC}) into (\ref{eq:aa}) yields:
\begin{align}
\dot{E}(S_\lambda)&= \frac{E(S_\lambda)}{2A(S_\lambda)}\int_{S_\lambda} \theta_+\,\mathrm{d}S_\lambda\nonumber\\
&+\frac{\sqrt{A(S_\lambda)}}{(4\pi)^{3/2}}\int_{S_\lambda}\bigg\{ -\theta_-\left(\frac{1}{2}\theta_+^2+\sigma_{ab}\sigma^{ab}+R_{ab}l^al^b\right)\nonumber\\
&+\theta_+\left(D_a\Omega^a+\Omega_a\Omega^a-\frac{1}{2}{}^2R+8\pi T_{ab}l^an^b   \right) \bigg\}\,\mathrm{d}S_\lambda\quad.
\label{eq:Edot}
\end{align}
Eardley \cite{Eardley:1979dra} established a monotonicity results for a particular family of surfaces $(S_r)$. Starting off with a surface $S$ with $\theta_+> 0$ \& $\theta_-\le 0$ almost everywhere. One can define a constant $r$ on $S$ by $A(S)=:4\pi r^2$. Although $n^a$ is normalised such that $n^al_a=-1$, there is still a pointwise rescaling freedom of $l^a$ left: $l^a\rightarrow \alpha l^a$ with $\alpha>0$. It is used to rescale $l^a$ such that $\theta_+=\frac{2}{r}$. Since $\partial_\lambda r=1$, $r$ is also a parameter along the congruence. In fact, $r$ corresponds to an area distance function
\begin{equation}
r=\sqrt{\frac{A}{4\pi}}\quad,
\end{equation}
which is related to the luminosity distance via Etherington's reciprocity theorem \cite{Etherington}.
Starting with the initial surface $S$ being a lightcone section arbitrarily close to the tip $p$, the remaining lightcone is foliated by level surfaces $S_r$ of constant $r$. Along this special family of surfaces $S_r$, (\ref{eq:Edot}) can be further simplified by inserting the explicit expressions for $A$ and $\theta_+$:
\begin{align}
\dot{E}(S_r)=\frac{1}{4\pi}\int_{S_r}\bigg\{ -\frac{r}{4}\theta_-\left(\sigma_{ab}\sigma^{ab}+R_{ab}l^al^b\right)\nonumber\\
+\frac{1}{2}\left(\Omega_a\Omega^a+\frac{1}{2}R+R_{ab}l^an^b\right) \bigg\}\,\mathrm{d}S_r\quad,
\label{eq:ss}
\end{align}
where the Gauss-Bonnet theorem for a sphere $\int_S{}^2R\,\mathrm{d}S=8\pi$ was used as well as $\int_S D_a\Omega^a\,\mathrm{d}S=0$, because $S$ is a closed surface. Extending the assumption $\theta_+>0$ and $\theta_-\le 0$ to all $S_r$, and further assuming the DEC, we find the right hand side of (\ref{eq:ss}) to be non-negative, because $\sigma_{ab}\sigma^{ab}\ge 0$ and $\Omega_a\Omega^a\ge 0$, which can be verified by direct calculation. Using the EFEs, the curvature terms are shown to be non-negative because of the DEC. Thus, the Hawking energy increases monotonously along the particular foliation $(S_r)$ of $C^-(p)$ given the above assumptions. In fact, monotonicity can be established for a whole class of foliations, namely those with $\partial_{\tilde{\lambda}}\lambda>0$, cf. (\ref{eq:ooo}). Eardley's precise expression in Newman-Penrose variables \cite{Eardley:1979dra} is recovered after making use of the identities $\mu=\frac{\theta_-}{2}$, $\frac{1}{2}\sigma_{ab}\sigma^{ab}=|\sigma|^2$, $\phi_{00}=\frac{1}{2}R_{ab}l^al^b$, $\Omega_a\Omega^a=2\pi\bar{\pi}=2|\alpha+\bar{\beta}|^2$ and $\frac{1}{4}R+\frac{1}{2}R_{ab}l^an^b=3\Lambda+\phi_{11}$.
Additionally, if the initial sphere $S$ is sufficiently close to the lightcone tip $p$, the Hawking energy is positive due to the small sphere limit of \cite{Horowitz:1982}. Summarising, we found that the Hawking energy on the past lightcone $C^-(p)$ of an observer $p$ is positive and monotonously increasing to the past, provided that $\theta_+>0$ \& $\theta_-\le 0$ almost everywhere, and matter obeys the DEC.

The central assumption is the strict positivity of the expansion $\theta_+$ of the outgoing null congruence $l^a$ generating the lightcone. Firstly, this excludes spacetimes with certain global properties, such as the existence of past apparent horizons, beyond which $\theta_+$ turns negative \cite{Ellis:2015928}. This is the case in many cosmological settings, in particular FLRW dust universes with a positive cosmological constant, the Einstein static universe or other recollapsing models. In any case, the monotonicity results remain true even in such spacetimes in a suitably close neighbourhood of the observer $p$. Secondly, this also excludes local regions of $C^-(p)$ with negative expansion, as is the case in the presence of caustics due to local inhomogeneities causing strong gravitational lensing. A discussion on the inclusion of caustics can be found in the next chapter, the results so far only hold in the case of an empty cut locus of $C^-(p)$, $L^-(p)=\emptyset$. In particular, this includes the weak lensing regime.

\section{Geometry of $C^-(p)$ in the presence of strong lenses}
\label{sec:stronglensing}
In more realistic cosmological set-ups, the past lightcone will typically display self-intersections. Considering that gravitational lenses, such as galaxy clusters, galaxies, or individual stars, exist on different scales, caustics are expected to from hierarchical patterns on the past lightcone. Ellis et al.\ \cite{Ellis:1998ha} estimated the total number of caustics on our past lightcone due to inhomogeneities to be of the order $10^{22}$. The presence of caustics was neglected in the original observational cosmology programme of reconstructing the spacetime metric and energy momentum tensor from observables \cite{Ellis1985315}. However, their presence affects cosmological distances in such a way that observed area distances to objects are increased \cite{Ellis:1998ha}. 

In general, these self-intersections may be arbitrarily complicated. Yet, it was shown that the multitude of these self-intersections can be divided into stable and unstable ones in the following sense. The set of points in spacetime $M$ that can be reached by the outgoing, respectively ingoing, null geodesic congruence emanating from an orientable, spacelike, smooth surface $S$ in $M$ is called wavefront, see e.g.\ \cite{Perlick2004}. The caustic of a wavefront is defined to be the set of points where the wavefront fails to be an immersed submanifold of $M$. In particular, the past lightcone $C^-(p)$ of $p$ is a wavefront if $S$ is chosen suitably close to $p$. Stability refers to arbitrarily small perturbations of the initial surface $S$, see e.g.\ \cite{Hasse_1996} for more details. A classification of stable caustics of wavefronts was established by \cite{Friedrich:1983vi,Low:1993,Hasse_1996,Low:1998}, using Arnol'd's singularity theory of Lagrangian and Legendrian maps \cite{arnold1985singularities,Ehlers2}. Of particular relevance for the present work, Low showed \cite{Low:1993,Low:1998} that only two types of stable caustics appear in the intersection of a lightcone with a spacelike hypersurface, referred to as cusp and swallow-tail singularities, cf.\ Fig.\ref{fig:swallowtail}. For a detailed account on the different types of caustics in the context of gravitational lensing, we refer the reader to \cite{Petters2001}.

In particular, we are interested the simple lensing configuration displayed in Fig.\ref{fig:spherlens}, consisting of a single swallow-tail singular point and two cusp ridges present in $C^-(p)\cap\Sigma_t$. The presence of self-intersections renders $C^-(p)$ Lipschitz continuous on the measure zero set of self-intersections, in other words, the light cone remains smooth almost everywhere.  The following analysis also extends to more general lensing configurations, in particular, the results may be applied to multi-lens configurations in which the single lens in Fig.\ref{fig:spherlens} occurs multiple times; see also the discussion in section \ref{sec:monotonicity}. More formally, the analysis given below includes the subclass of configurations in which the singular points as well as the cut loci in $S$ are isolated and have measure zero. Physically, these configurations may be thought of as multiple local overdensities leading to isolated gravitational lenses of the type shown in Fig.\ref{fig:spherlens}.

Since singular points are conjugate points along the null generators with respect to $p$, the expansion of the lightcone generators $\theta_+=-\infty$ at a singular point. Hence, $\theta_+$ is a smooth function almost everywhere on $S$, apart from the singular points. Large regions of $S$ will display a positive $\theta_+$, and by continuity, any singular point is surrounded by a neighbourhood with negative $\theta_+$.

Also, at least for the lens configuration in Fig.\ref{fig:spherlens} and multi-lens set-ups thereof, only swallow-tail singular points can be found in the exterior part $C^-(p)\cap \dot{I}^-(p)$, whereas cusp singular points exclusively appear at self-intersections in the interior $I^-(p)$. Hence, for the configurations considered in this paper, it suffices to take only swallow-tail singular points on $S$ into account. Moreover, these observations seem to suggest that of all stable singular points according to Arnol'd, only swallow-tail ones appear in exterior part $C^-(p)\cap \dot{I}^-(p)\cap \Sigma_t$, however, it would be desireable to make a more rigorous statement on this matter.

\section{Hawking energy in the presence of singularities}
\label{sec:singularities}
Given the above set-up of a light cone including caustics, it is a natural question whether or not the Hawking energy for a surface containing these types of singularities is well-defined. In particular, since at singular points $\theta_+=-\infty$, one might wonder whether the integral $\int_S \theta_+\theta_-\,\mathrm{d}S$ in (\ref{eq:energy}) is well-defined, that is finite. In the following, we show that this is indeed the case for swallow-tail singularities only, whereas the integral is divergent if $S$ contains cusp points. Therefore, a finite Hawking energy can only be assigned to the exterior part of the lightcone, since the presence of cusp singularities in the interior causes the energy to diverge when integrated over the whole lightcone slice.

\begin{figure}
	\includegraphics[width=0.4\textwidth]{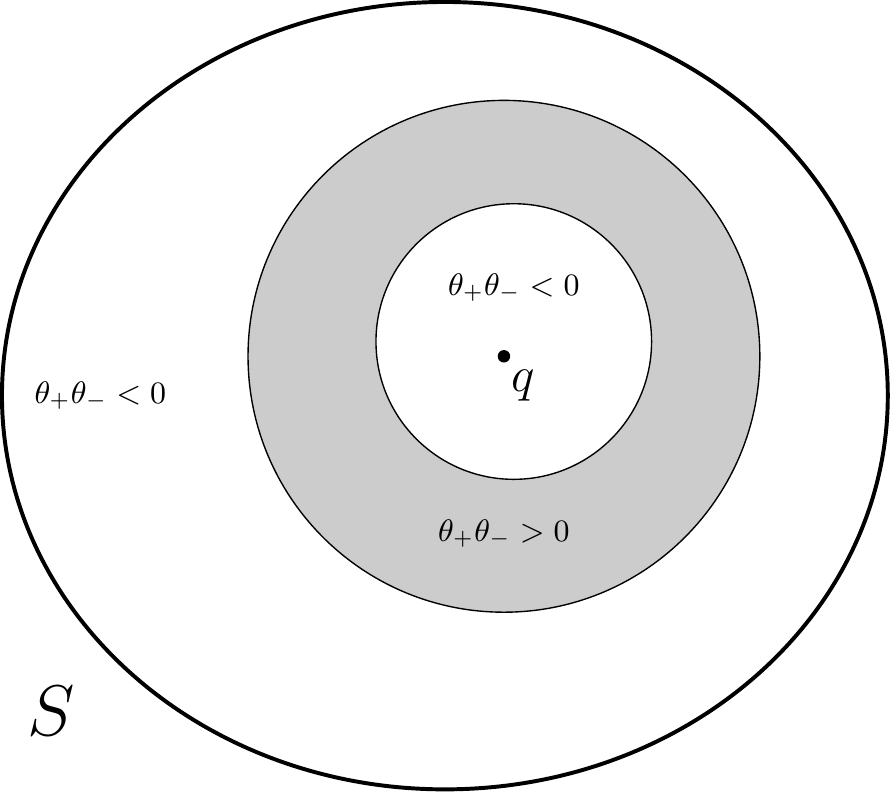}
	\caption{A singular point $q$ on $S$ is surrounded by a trapped, ring-like region (grey), where the product of the null expansions $\theta_+\theta_-$ is positive.}
	\label{fig:trapped ring}
\end{figure}

The past-pointing null vectors $l^a$ and $n^a$ orthogonal to the spacelike codimension-2 surface $S$ can be decomposed into a timelike (future-pointing) unit normal $t^a$ as well as an orthogonal spacelike unit normal $v^a$ via
\begin{equation}
l^a=\frac{1}{\sqrt{2}}\left(-t^a+v^a\right)\quad\&\quad 
n^a=\frac{1}{\sqrt{2}}\left(-t^a-v^a\right)\quad.
\label{eq:ln}
\end{equation}
Following \cite{Szabados:2004vb}, the tangent bundle $TM$ of the spacetime $M$ can be decomposed into the sum of 
the tangent bundle $TS$ of $S$ and the normal bundle $NS$ of $S$ by using the corresponding projectors
\begin{align}
\Pi^a_b&:=\delta^a_b+t^at_b-v^av_b=\delta^a_b+l^an_b+n^al_b\quad\text{and}\nonumber\\
O^a_b&:=\delta^a_b-\Pi^a_b
\end{align}
respectively. For example, the spacetime metric $g$ can be decomposed into the intrinsic metric $h$ on $S$ and an orthogonal part, cf.\ (\ref{eq:hab}):
\begin{equation}
g_{ab}=h_{ab}-t_at_b+r_ar_b=h_{ab}-l_an_b-n_al_b\quad.
\label{eq:metricdecomp}
\end{equation}
Corresponding to each normal, there is an associated extrinsic curvature $\tau_{ab}$ and $H_{ab}$:
\begin{equation}
\tau_{ab}=\Pi^c_a\Pi^d_b\nabla_ct_d \quad\&\quad H_{ab}=\Pi^c_a\Pi^d_b\nabla_cv_d\quad.
\label{eq:extrinsiscurv}
\end{equation}
$\tau_{ab}$ is the extrinsic curvature (or second fundamental form) of the spacelike hypersurface $\Sigma$ with timelike normal $t^a$ embedded in spacetime. $\Sigma$ creates the lightcone slices: $S=C^-(p)\cap\dot{I}^-(p)\cap \Sigma$. $H_{ab}$ is the extrinsic curvature of the spacelike 2-surface $S$ with spacelike normal $v^a$ within the spacelike hypersurface $\Sigma$. Taking the trace results in the mean curvatures $\tau$ and $H$ of $S$ in each direction. Using the definitions $\theta_+=\nabla_al^a$ and $\theta_-=\nabla_an^a$, together with (\ref{eq:ln}) \& (\ref{eq:extrinsiscurv}), yields the following relation between the null expansions and mean curvatures:
\begin{equation}
\theta_\pm= \frac{1}{\sqrt{2}}\left(-\tau\pm H\right).
\label{eq:expansions}
\end{equation}
Because $\theta_+=-\infty$ at singular points, (\ref{eq:expansions}) implies that one of the mean curvatures has to diverge. Since $\tau$ is the mean curvature of the smooth spacelike hypersurface $\Sigma$, it is finite, and hence, $H\rightarrow-\infty$ at singular points. The product of expansions can be written as the norm of the main curvature vector $Q^a$ of $S$:
\begin{align}
Q^a:&=-\theta_-l^a-\theta_+ n^a=\tau t^a-H v^a\quad\text{thus}\nonumber\\
-Q^aQ_a&=2\theta_+\theta_-=\tau^2-H^2\quad.\label{eq:meancurv}
\end{align}
In the generic case where $\theta_+>0$ and $\theta_-<0$, $Q^a$ is spacelike, it is null if one of the expansions is zero, for example on horizons, and becomes timelike if the expansions have the same sign, for instance for trapped surfaces. These results imply that every singular point on $S$ is surrounded by a "trapped ring", where $\theta_+\theta_->0$ (see Fig.\ref{fig:trapped ring}).

Using (\ref{eq:meancurv}), the integral expression appearing in (\ref{eq:energy}) becomes
\begin{equation}
E(S)=\frac{\sqrt{A(S)}}{(4\pi)^{3/2}}\left(2\pi+\frac{1}{8}\int_S \left(\tau^2-H^2\right)\,\mathrm{d}S\right)\quad.
\label{eq:ww}
\end{equation}

The fact that $H$ diverges at singular points was also more formally established in \cite{geometryoffronts} (c.f.\ corollary 3.5), where the authors proved that 
the mean curvature of a hypersurface in a Riemannian manifold diverges at swallow-tail or cusp singular points. Although $H$ is divergent, one might still hope that $\int_S H^2\;\mathrm{d}S$ in (\ref{eq:ww}) is finite. In the following, we show that this is the case only for swallow-tail singularities, whereas the integral diverges for cusp singularities. Therefore, the Hawking energy is well-defined, i.e.\ finite-valued, for Lipschitz surfaces only containing swallow-tail singularities. 

It suffices to study the integral in a neighbourhood $Q\subset S$ of a singular point, because the mean curvature is always finite-valued at non-singular points of $S$. Below, we arrive at explicit expressions for the mean curvature near a cusp and swallow-tail singular point, if $S$ is embedded in Euclidean space, and find that $\int_Q H^2\;\mathrm{d}S$ diverges for a cusp, but is finite for a swallow-tail point. This result can be immediately extended to the Riemannian case by replacing the Euclidean metric $g$ in the calculation below with its Riemannian counterpart, altering the result only by finite factors.

\begin{figure}
	\includegraphics[width=0.49\textwidth]{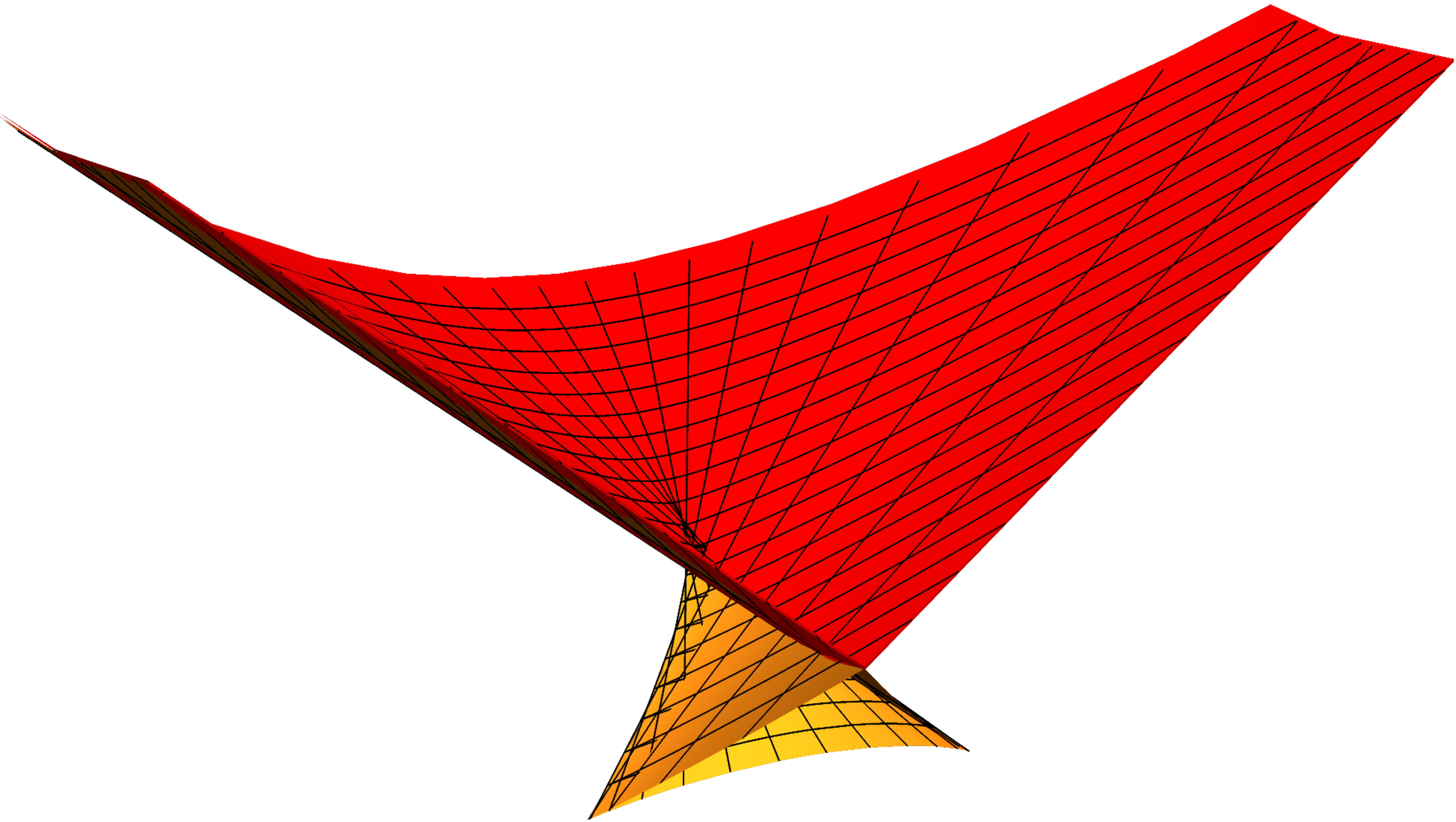}
	\caption{Front containing a swallow-tail singularity. The outer part $\subset \dot{I}^-(p)$ is coloured in red and satisfies $v\geq-2u^2$. It is a zoom-in of Fig.\ref{fig:spherlens} near the swallow-tail point $q$.}
	\label{fig:swallowtail}
\end{figure}

The following calculation and notation follows \cite{geometryoffronts}. Given a smooth map  $f: M\rightarrow N$ from an oriented 2-manifold $M$ into an oriented Riemannian 3-manifold $N$ with metric $g$. $f$ is called an instantaneous wavefront if there exists a unit vector field $\nu^a\in N$ along $f$ such that $g(f_*X,\nu)=0\quad \forall X\in TM$. $\nu^a$ is called the normal vector of the instantaneous wavefront $f$. An instantaneous wavefront is the intersection of a wavefront with a spacelike hypersurface \cite{Perlick2004}. $q\in M$ is called a singular point of the front $f$, if $f$ is not an immersion at $q$. A singular point is called cusp point or swallow-tail point respectively, if it is locally diffeomorphic to 
\begin{align}
f_C(u,v)&:=(u^2,u^3,v)\quad\text{or}\nonumber\\ f_S(u,v)&:=(3u^4+u^2v,4u^3+2uv,v)\label{eq:singularities}
\end{align}
at $(u,v)=(0,0)$. The mean curvature $H$ of the front $f$ with normal vector $\nu^a$ is
\begin{equation}
H:=\frac{EN-2FM+GL}{4\lambda^2}\quad,
\end{equation}
with $f_u=\partial_uf$, $f_v=\partial_vf$, $E=g(f_u,f_u)$, $F=g(f_u,f_v)$, $G=g(f_v,f_v)$, $|\lambda|=\sqrt{EG-F^2}$, $L=-g(f_u,\nu_u)$, $M=-g(f_v,\nu_u)=-g(f_u,\nu_v)$, $N=-g(f_v,\nu_v)$.
Computing the mean curvature for cusp and swallow-tail singularities near the singular point $(0,0)$ yields:
\begin{align}
H_C&=-\frac{3}{2u(9u^2+4)^{3/2}}\quad\text{and}\\ H_S&=\frac{u^4+4u^2+1}{8(6u^2+v)(u^4+u^2+1)^{3/2}}\quad.
\label{eq:meancurvatures}
\end{align}
Inserting these into the integral expression in (\ref{eq:ww}) using $\mathrm{d}S=|\lambda|\,\mathrm{d}u\,\mathrm{d}v$ and setting the integration range\\ $Q_C=\left\{v\in[b_1,b_2],u\in[-a,a]\right\}$ yields
\begin{align}
&\int_{Q_C} H_C^2\,\mathrm{d}S= \frac{9}{4}v|_{b_1}^{b_2}\int_{-a}^a\frac{\mathrm{d}u}{|u|\left(9u^2+4\right)^{5/2}}\nonumber\\&\quad\overset{u\rightarrow 0}{\approx} \frac{9}{128}v|_{b_1}^{b_2}\int_{-a}^a\frac{\mathrm{d}u}{|u|}
=\frac{9}{128}(b_2-b_1)\cdot 2\left[\ln(|u|)\right]_0^a=+\infty
\end{align}
for the cusp case. In the case of the swallow-tail, we must be careful to only integrate over the outer part of the surface, i.e.\ the part contained in $C^-(p)\cap\dot{I}^-(p)$, see Fig.\ref{fig:swallowtail}. This is done by restricting the integration range to $v\geq-2u^2$ in the above parametrization, \\$Q_S=\left\{ v\in[-2u^2,a], u\in[-b,b]  \right\}$, yielding
\begin{align}
\int_{Q_S} H_S^2\,\mathrm{d}S&=\int_{-b}^{b}\mathrm{d}u\int_{-2u^2}^{a}\mathrm{d}v\, \frac{\left(u^4+4u^2+1\right)^2}{32\left( u^4+u^2+1\right)^{5/2}(6u^2+v)}\nonumber\\
&= \int_{-b}^b\mathrm{d}u\, \frac{\left(u^4+4u^2+1\right)^2}{32\left( u^4+u^2+1\right)^{5/2}}\ln\left(\frac{3}{2}+\frac{a}{4u^2}\right)\nonumber\\
&\overset{u\rightarrow 0}{\approx}\frac{1}{32} \int_{-b}^{b}\ln\left(\frac{a}{4u^2}\right)\,\mathrm{d}u\nonumber\\
&=\frac{1}{16}\left[2u+u\ln\left(\frac{a}{4u^2}\right)\right]_0^b<\infty\quad.
\end{align}
Summarising, we showed that the integrals in (\ref{eq:ww}) are finite, thus the Hawking energy for a topological sphere $S$ containing swallow-tail singularities is well-defined.

Next, we address the derivative of $E$ along the null generators (\ref{eq:Edot}), which can be further simplified by making use of the contracted Gauss equation, cf. \cite{Gourgoulhon:2005ng}:
\begin{equation}
{}^2R=h^{ac}h^{bd}R_{abcd}-\theta_+\theta_-+2\sigma^+_{ab}\sigma_-^{ab}\quad.
\end{equation}
Applying the Ricci decomposition of the Riemann tensor (see also \cite{Faraoni:2015cnd}) and using the metric decomposition (\ref{eq:metricdecomp}) together with the EFE yields:
\begin{align}
h^{ac}h^{bd}R_{abcd}&=h^{ac}h^{bd}C_{abcd}+16\pi T_{ab}l^an^b+\frac{8\pi}{3}T\quad\text{with}\nonumber\\
h^{ac}h^{bd}C_{abcd}&=2C_{lnnl}\quad,
\end{align}
after using the symmetries of the Weyl tensor. The Weyl tensor term vanishes if $l^a$ belongs to a null geodesic congruence. This can be seen by taking the shear evolution equation along the null congruence, contracting it with $n^an^b$ and noting that $\sigma^+_{ab}n^a=0$:
\begin{equation}
C_{nlnl}=-n^an^b\nabla_l\sigma^+_{ab}-\theta_+\sigma^+_{ab}n^an^b=0\quad.
\end{equation}
Summarising, we find
\begin{equation}
{}^2R=16\pi T_{ln}+\frac{8\pi}{3}T-\theta_+\theta_-+2\sigma^+_{ab}\sigma_-^{ab}\quad.
\label{eq:2dricci}
\end{equation}

Next, the term $\Omega_a\Omega^a$ appearing in (\ref{eq:Edot}) can be expressed in terms of the energy momentum tensor. \\Recalling that $\Omega_a=\nabla_ln_a$ as well as taking into account that $\Omega_al^a=0=\Omega_an^a$, we are left with
\begin{align}
\nabla_l\left( n^\alpha\nabla_ln_\alpha \right)&=\nabla_l\left(\frac{1}{2}\nabla_l(n_\alpha n^\alpha)\right)=0\qquad\Leftrightarrow\nonumber\\
\Omega_a\Omega^a&=-n^a\nabla_l\Omega_a\quad.
\end{align}
Using the evolution equation for $\Omega_a$ along the null \\generators (cf.\ \cite{Gourgoulhon:2005ng}),
\begin{equation}
\nabla_l\Omega_\nu=-\Theta_\nu^{\,\alpha} \Omega_\alpha-\theta_+\Omega_\nu+8\pi T_{\mu\nu}l^\mu+\frac{1}{2}D_\nu\theta_+-D_\alpha\sigma_{+\;\nu}^{\,\alpha}\quad,
\end{equation}
and contracting it with $n^a$, we end up with
\begin{equation}
\Omega_a\Omega^a=-8\pi T_{ln}\quad.
\label{eq:omega}
\end{equation}
Thus, inserting (\ref{eq:2dricci}) and (\ref{eq:omega}) into (\ref{eq:Edot}), we find
\begin{widetext}
\begin{equation}
\dot{E}=\frac{E(S)}{2A(S)}\int_S \theta_+\,\mathrm{d}S
+\frac{\sqrt{A(S)}}{(4\pi)^{3/2}}
\int_S \left\{-\left( \theta_-\sigma^+_{ab}\sigma_+^{ab}+\theta_+\sigma^+_{ab}\sigma_-^{ab} \right)
-8\pi\left(\theta_-T_{ll}+\theta_+\left[T_{ln}+\frac{1}{6}T\right]\right)+\theta_+D_\alpha\Omega^\alpha\right\}\,\mathrm{d}S\quad.
\label{eq:XXX}
\end{equation}
\end{widetext}
This expression describes how the Hawking energy changes along constant affine parameter slices of spherical topology of the past lightcone $C^-(p)$. The shear and matter effects separate into two different contributions.

Before discussing the terms and addressing monotonicity, we first comment on whether or not the first derivative $\dot{E}$, in addition to the energy itself, is well-defined. One can check with (\ref{eq:meancurvatures}) that $\int_S H_S\,\mathrm{d}S$ as well as $\int_S H_S^2\,\mathrm{d}S$ are finite, however, $\int_S H_S^3\,\mathrm{d}S$ diverges. Since the first as well as the terms involving the energy momentum tensor are proportional to $H$, they are finite. Because $S$ is a manifold without boundary, $\int_S \theta_+D_\alpha\Omega^\alpha=-\int_S \Omega^\alpha D_\alpha\theta_+$, and $D_\alpha\theta_+$ is proportional to $\partial_uH$ and $\partial_vH$. Again, one can check explicitly with the help of (\ref{eq:meancurvatures}) that these derivatives are finite. Turning to the shear terms, we note that the shear tensors $\sigma^\pm_{ab}$ are also diverging at singular points, in particular in the same way as $\theta_\pm$ for non-degenerate singular points such as cusp or swallow-tail \cite{Seitz:1994xf}. Therefore, $\theta_-\sigma^+_{ab}\sigma_+^{ab}$ and $\theta_+\sigma^+_{ab}\sigma_-^{ab}$ are of order $H^3$, hence their integrals over $S$ diverge. However, close to a singular point, both terms have opposite signs and cancel each other. This can be seen by rewriting the expression with the help of $Q_{ab}^c:=h^d_a h^e_b\nabla_d h^c_e$ and noting that $\Theta_{ab}=-l_cQ_{ab}^c$, $\Xi_{ab}=-n_c Q_{ab}^c$:
\begin{equation}
\theta_-\sigma^+_{ab}\sigma_+^{ab}+\theta_+\sigma^+_{ab}\sigma_-^{ab}=\sigma_+^{ab}Q_cQ_{ab}^c\quad.
\end{equation}
Next, expressing $Q^a$ and $Q_{ab}^C$ in terms of the timelike and spacelike unit normals $t^a$ and $v^a$ yields:
\begin{align}
\sigma_+^{ab}Q_c&Q_{ab}^c= \frac{1}{\sqrt{2}}\tau\tau_{ab}\left(\tau^{ab}+H^{ab}\right)+\frac{1}{2}\tau^2\theta_+\nonumber
\\&-\frac{1}{\sqrt{2}}HH_{ab}\tau^{ab}
+\frac{1}{\sqrt{2}}HH_{ab}H^{ab}-\frac{1}{2}\theta_+H^2\quad.
\end{align}
All terms apart from the last two are at most of the order $H^2$ and thus integrable. The last two terms are diverging as $H^3$ near a singular point, but being of opposite sign, they precisely cancel each other. Hence, the integral of the shear terms is also finite.

Summarising, we found that the Hawking energy as well as its first derivative along the null generators of the past lightcone are well-defined even for surfaces including swallow-tail type singularities. Knowing that (\ref{eq:XXX}) is a well-defined quantity, we now address the rescaling freedom of $l^a$ before studying the monotonicity of (\ref{eq:XXX}).

\section{Choice of rescaling}
\label{sec:rescaling}
As mentioned earlier, once a scaling function $\alpha$ is chosen, the Hawking energy will monotonously increase and be positive along the family of constant (affine) parameter surfaces $(S_\lambda)$ associated with this rescaling, if and only if (\ref{eq:XXX}) is positive. Hayward \cite{Hayward:1993ma} pointed out that the sign of $\int_S \theta_\pm\,\mathrm{d}S$ is not an invariant under rescaling $l^a$. In particular, if $\theta_\pm$ changes its sign on $S$, $\int_S \theta_\pm\,\mathrm{d}S$ can take any sign and value by constructing an appropriate rescaling function $\alpha$ on $S$. In fact, since all terms appearing in (\ref{eq:XXX}) are not invariant under rescaling, one can use the rescaling freedom to simplify its right-hand-side. As in the weak lensing case, the term $\int_S\theta_+D_a\Omega^a$ can be eliminated even if $\theta_+$ is not strictly positive anymore. Under rescaling $l^a\rightarrow\alpha l^a$, $\alpha>0$, $\Omega_a$ transforms as
\begin{equation}
\Omega_a\rightarrow\Omega+D_a\ln\alpha\quad\Rightarrow\quad D_a\Omega^a\rightarrow D_a\Omega^a+D_aD^a\ln\alpha\quad,
\end{equation}
leading to the following Poisson equation for $\alpha$ on $S$, if the rescaling is used to eliminate $D_a\Omega^a$:
\begin{equation}
D_aD^a\ln\alpha=-D_a\Omega^a\quad.
\label{eq:poisson}
\end{equation}
We have two cases to consider depending on whether $S$ is smooth or Lipschitz.

\subsection{Poisson equation on smooth Riemannian manifold}
For smooth $S$, we have the following existence theorem for the Poisson equation:\\

\textit{Theorem:} On a closed Riemannian manifold $M$, if $\rho$ is a smooth function satisfying $\int_M \rho=0$: $\exists$ smooth solution to $\Delta\Phi=\rho$, unique up to the addition of a constant.\\

Since $S$ is a closed manifold, $\int_SD_a\Omega^a=0$ and therefore we can find an $\alpha$ such that $D_a\Omega^a=0$ after rescaling.

\subsection{Poisson equation on a Lipschitz manifold}
If $S$ is only Lipschitz continuous, we would still like to eliminate $\int_S\theta_+D_a\Omega^a$. Because the Poisson equation (\ref{eq:poisson}) contains second derivatives, it is ill-defined on a Lipschitz manifold. However, one can adapt a weak (i.e.\ distributional) formulation in the following way. Recall that a Riemannian Lipschitz manifold $(M,q)$ is a manifold $M$ equipped with a positive-definite metric $q$, for which all transition maps are locally Lipschitz functions. By Rademacher's theorem, a locally Lipschitz function $f:\mathbb{R}^n\rightarrow\mathbb{R}^m$ is differentiable almost everywhere (w.r.t\ to the n-dim. Lebesque measure). We denote the linear space of all Lipschitz continuous functions $\phi: M\rightarrow\mathbb{R}$ for which the norm
\begin{equation}
||\phi||^2:=\int_M (\phi^2+|\nabla\phi|^2)d\mu\;<\infty
\end{equation}
is finite by $\text{Lip}^{1,2}(M)$. This norm is well-defined because Rademacher's theorem ensures the existence of the gradient almost everywhere. We then define the Sobolev space $W^{1,2}(M)$ as the Cauchy completion of $\text{Lip}^{1,2}(M)$ with respect to the above norm $||\,.\,||$ . If $M$ is a compact, connected, oriented, Lipschitz manifold without boundary, the weak version of the Poisson equation reads
\begin{equation}
-\int_M\braket{\nabla\phi,\nabla f}d\mu=\int_M g\, \phi\,d\mu\quad,
\end{equation}
$\forall\, \phi\in W^{1,2}(M)$, given that $f\in W^{1,2}(M)$ and $g\in L^2(M)$ (see theorem 1.3 in \cite{laplacebeltrami}). Thus, provided that $\theta_+\in W^{1,2}(M)$ and $D_a\Omega^a\in L^2(M)$, we can find a function $\alpha$ such that 
\begin{equation}
\int_S \theta_+\left(D_a\Omega^a+D_aD^a\ln\alpha\right)\mathrm{d}S=0\quad.
\end{equation}

Hence, one way to use the rescaling freedom is to eliminate the term $\int_S\theta_+ D_a\Omega^a\,\mathrm{d}S$ in (\ref{eq:XXX}), provided that $\theta_+\in W^{1,2}(S)$. $D_a\Omega^a$ is in $L^2(M)$ because of $\int_S D_a\Omega^a\,\mathrm{d}S=0$. If $\theta_+\notin W^{1,2}(S)$, we can use the rescaling of $l^a$ to achieve $\int_S\theta_+\,\mathrm{d}S>0$, but then assumptions on $D_a\Omega^a$ have to be made.

\section{Monotonicity}
\label{sec:monotonicity}
The crucial difference to the weak lensing case is that there now exists a region of negative $\theta_+$ on $S$ connected to the singular point. This implies that locally, the area decreases along $l^a$ although the total area of $S$ can still increase if $\dot{A}(S)=\int_S\theta_+\,\mathrm{d}S>0$. It is precisely this region in which energy can now be injected into the interior of $\dot{I}^-(p)$ from the exterior along causal curves. Hence, the naive monotonicity argument related to Fig.\ref{fig:monotonicity} holds only for regions with positive $\theta_+$ and fails in regions of negative $\theta_+$. In general, two different effects concerning monotonicity have to be taken into account:
\begin{itemize}
	\item[(i)] A variation in the area $A$ leads to a change in the energy, because the amount of matter enclosed by $S$ changes. This effect is manifested in the first term in (\ref{eq:XXX}), describing nothing other than the change of $A$ along the null generators $l^a$.
	\item[(ii)] Energy may leave $I^-(p)$ only in regions with $\theta_+>0$, and enter $I^-(p)$ only where $\theta_+<0$. This is accounted for by the second integral in (\ref{eq:XXX}), stating two contributions: shear and matter. The first corresponds to energy transported by the pure gravitational field in the form of gravitational waves, and is even present in vacuum. The latter contribution is due to matter encoded in the energy momentum tensor satisfying the DEC. A potential cosmological constant can be accommodated in the energy momentum tensor.
\end{itemize}
Assume in the following that $A$ is increasing along the family of surfaces, i.e. $\int_S \theta_+\,\mathrm{d}S>0$. In the case of a vacuum spacetime, the matter terms vanish and one only has to deal with the net flux of in- and outgoing shear contributions. Furthermore, by the Goldberg-Sachs theorem \cite{Goldberg2009}, the geodesic congruence $l^a$ in a vacuum spacetime $M$ is shear free, i.e.\ $\sigma^+_{ab}=0$, if and only if $M$ is algebraically special, that is $l^a$ is a repeated principle null direction, see also \cite{Ellis:2011pi} and references therein for generalisations. Demanding vanishing shear within the class of non-vacuum spacetimes imposes a strong constraint, see \cite{Adamo2012,Ellis:2011pi}.\\

So far, the studied configuration contained only one strong gravitational lens, see Fig. \ref{fig:spherlens}. Nevertheless, the obtained results can easily be generalised to configurations with multiple isolated strong lensing events taking place, that is, the swallow-tail singular points on $S$ have to be isolated. The more lensing events happen, the larger the fraction of $S$ with negative $\theta_+$. Having more and more lensing events present will ultimately turn $\int_S\theta_+\,\mathrm{d}S$ negative and therefore the whole lightcone will refocus. This indicates that enough energy is concentrated in the interior to cause the shrinking of $S$.

\section{Extensions}
\label{sec:extensions}
Until now, the discussion was restricted to a family of topological 2-spheres. In the following, we briefly review how a change of topology affects the results. One could imagine that more complicated lensing configurations may cause $C^-(p)\cap\dot{I}^-(p)$ to still consist of one connected component, but to be topologically different from $\mathbb{R}\times S^2$. The Hawking energy can be generalised to arbitrary closed orientable surfaces $\tilde{S}$, characterised by their genus $g$, by using the Gauss Bonnet theorem $\int_{\tilde{S}} {}^2R\,\mathrm{d}\tilde{S}=8\pi(1-g)$, see for instance \cite{Hayward:1993ph}. Then, (\ref{eq:energy}) reads instead
\begin{equation}
E(\tilde{S}):= \frac{\sqrt{A(\tilde{S})}}{32\pi^{3/2}}\left(8\pi(1-g(\tilde{S}))+\int_{\tilde{S}}\theta_+\theta_-\,\mathrm{d}\tilde{S}\right)\quad.
\end{equation}
However, if $g(\tilde{S})\geq 1$, and the surface is non-trapped on average in the sense of \cite{Hayward:1993ma}, i.e.\ $\int_{\tilde{S}} \theta_+\theta_-\,\mathrm{d}\tilde{S}<0$, the Hawking energy is negative.

In principle, it is also possible that $C^-(p)\cap\dot{I}^-(p)$ splits into $n$ disconnected components of genus $g_i$ : 
\begin{equation}
C^-(p)\cap\dot{I}^-(p)=\mathbb{R}\times S_{g_1}\times\dots \times S_{g_n}\quad.
\end{equation}
This can happen for instance in situations topologically similar to a Schwarzschild black hole (see e.g.\ \cite{Perlick2004}). Hayward \cite{Hayward:1993ph} observed that the Hawking energy for $n$ disconnected surfaces is superadditive. If we denote $S_1\cup\dots\cup S_n=S_{\cup}$, then 
\begin{equation}
E_\cup=\sqrt{\frac{A_\cup}{A_1}}E_1+\dots+\sqrt{\frac{A_\cup}{A_n}} E_n>E_1+\dots+E_n\quad.
\end{equation}
This property is in contrast to the expected subadditivity of gravitational systems.

\section{Conclusions}
\label{sec:conclusions}
The Hawking energy provides a reasonable definition of energy in the setting of cosmology. The past lightcone of a point $p$ in spacetime is closely linked to cosmological observations and therefore provides the ideal geometric structure to study the properties of the Hawking energy in a physical set-up. The part of it within the causal boundary, $C^-(p)\cap\dot{I}^-(p)$, provides the mathematical arena in which the Hawking energy is studied. It is a Lipschitz continuous manifold, potentially containing points where $\theta_+$ is singular, and is assumed to have spherical topology: $C^-(p)\cap\dot{I}^-(p)\simeq\mathbb{R}\times S^2$. This seems to be a natural case, but it would be interesting to get a better understanding whether, and under what circumstances other topologies may arise. 

Assuming that the universe is described by a globally hyperbolic spacetime in which all matter obeys the DEC, strong gravitational fields may cause the lightcone to intersect itself locally at singular points, or globally. Since these singular points are conjugate to $p$, the presence of singularities indicates the existence of regions on $S$ where the expansion $\theta_+$ of the null generators is negative. The only two stable types of singularities appearing in lightcone slices are cusp and swallow-tail singularities \cite{Low:1993}. As a main result, we find that the Hawking energy of surfaces containing cusp singular points is divergent, but finite for swallow-tail singularities.

For a lightcone, the presence or absence of self-intersections gives rise to two natural regimes. The weak lensing regime, in which self-intersections are absent, exhibits a positive expansion parameter $\theta_+>0$ everywhere on smooth surfaces $S$. The Hawking energy is positive and monotonously increases along the null generators of the past lightcone, following directly from Eardley's results \cite{Eardley:1979dra} and the small sphere limit \cite{Horowitz:1982}. Studying a lightcone going through multiple isolated strong gravitational lenses, for instance caused by isolated local matter overdensities, we find that the Hawking energy associated with the exterior part of a lightcone slice, exclusively containing swallow-tail singular points, is finite. The energy of the total lightcone slice, however, is infinite due to the presence of cusp singularities in the interior part. Monotonicity (\ref{eq:XXX}) depends upon two effects. Firstly, the area of $S$ changes along the null generators. Secondly, and in contrast to the weak lensing case, matter may enter the interior of $\dot{I}^-(p)\cap C^-(p)$ through regions where $\theta_+<0$. Hence in general, the Hawking energy is not monotonous along the past lightcone anymore and monotonicity depends on the balance of in- and outgoing energy flux.

More generally, since the lightcone construction differs from other instantaneous wavefronts only by the small sphere limit, 
the results concerning well-definedness and monotonicity of the Hawking energy extend to all instantaneous wavefronts. In particular, wavefronts containing swallow-tail singular points have a finite associated Hawking energy, whereas the energy of wavefronts containing cusp singularities diverges. 

With the above results at hand, it might be interesting to study the energy of general wavefronts, such as light signals or gravitational waves, propagating through an (inhomogeneous) universe. In the particular context of inhomogeneous cosmology, it might be a useful tool to compare an inhomogeneous universe with a FLRW-reference universe and thereby address the so-called fitting problem in inhomogeneous cosmology \cite{ellis_maartens_maccallum_2012} by comparing domains of equal mass or energy, as it was advocated in \cite{Ellis:1998ha}.

\begin{acknowledgments}
	The author would like to thank L. Brunswic, T. Buchert, M. Carfora, D. Giulini, V. Perlick, and R. Tanzi for fruitful discussions and comments. This work is supported by the DFG Research Training Group "Models of Gravity" as well as by the ERC Advanced Grant 740021--ARTHUS, PI: T. Buchert.
\end{acknowledgments}

\interlinepenalty=10000
\enlargethispage{3\baselineskip}
\bibliography{HawkingEnergy}

\end{document}